\documentclass{elsart}

\usepackage{graphicx}
\usepackage[dvips,colorlinks=true,linkcolor=blue]{hyperref}
\usepackage{amssymb,amsmath,amsfonts,bm}

\begin{document}

\begin{frontmatter}



\title{Spiral modes in the diffusion of a granular particle on a vibrating surface}
\author[mpi,qmw]{R. Klages\corauthref{cor1},}
\ead{rklages@mpipks-dresden.mpg.de}
\ead[url]{www.mpipks-dresden.mpg.de/\~{}rklages}
\corauth[cor1]{corresponding author}
\author[mpi,tuv]{I.F. Barna,}
\ead{barna@mpipks-dresden.mpg.de}
\ead[url]{www.mpipks-dresden.mpg.de/\~{}barna}
\author[cenoli]{L. M\'aty\'as}
\ead{lmatyas@ulb.ac.be}
\address[mpi]{Max Planck Institute for the Physics of Complex Systems,
N\"othnitzer Str.\ 38, D-01187 Dresden, Germany}
\address[qmw]{School of Mathematical Sciences,
Queen Mary, University of London, Mile End Road, London E1 4NS, UK}
\address[tuv]{Institute for Theoretical Physics, TU Vienna, Wiener
Hauptstr.\ 8-10, A-1040 Vienna, Austria}
\address[cenoli]{Center for Nonlinear Phenomena and Complex Systems,
Universit\'{e} Libre de Bruxelles, Campus Plaine CP 231, Blvd.\ du Triomphe,
B-1050 Brussels, Belgium}

\begin{abstract}
We consider a particle that is subject to a constant force and scatters
inelastically on a vibrating periodically corrugated floor. At small friction
and small radius of the circular scatterers the dynamics is dominated by
resonances forming spiral structures in phase space. These spiral modes lead
to pronounced maxima and minima in the diffusion coefficient as a function of
the vibration frequency, as is shown in computer simulations. Our theoretical
predictions may be verified experimentally by studying transport of granular
matter on vibratory conveyors.
\end{abstract}


\end{frontmatter}

\section{Introduction}

A ball bouncing inelastically on an oscillating plate under the action of a
gravitational field appears to be a very simple dynamical
system. Surprisingly, such a ball exhibits an extremely rich dynamics thus
providing a prominent example for complexity in a seemingly trivial nonlinear
dynamical system: Experiments indicated period-doubling bifurcations into
chaotic motion \cite{Pier83,Pier85,KFP88,TAR92} while theoretical analyses
classified different types of dynamical behavior in terms of phase locking,
chattering, and strange attractors \cite{LuMe93,WCBH96,GiMa03}.

Oscillating surfaces, on the other hand, are often used in the field of
granular material in order to drive a gas of granular particles into a
nonequilibrium steady state \cite{LCG99,FSWV02,PrEgUr02}. A model that is
somewhat intermediate between an interacting many-particle system and a single
bouncing ball is the {\em bouncing ball billiard} (BBB)
\cite{MaKl03}. Here the surface is not flat but periodically corrugated
mimicking a fixed periodic grid of steel balls \cite{PrEgUr02}. A single
particle bouncing inelastically on this vibrating surface thus performs
diffusive motion.  Studying the BBB at large friction and large radius of the
scatterers, the diffusion coefficient turned out to be a highly irregular
function of the driving frequency \cite{MaKl03}. As the dynamical reason for
the largest maxima some integer frequency locking between bouncing ball and
vibrating plate could be identified, a phenomenon that is well-known for a
ball bouncing on a flat vibrating plate
\cite{Pier85,KFP88,TAR92,LuMe93,WCBH96,HCF89,HoFi89}. Additionally, we
detected an irregular structure on fine scales that is supposedly due to some
subtle effects like pruning of orbits under parameter variation \cite{KlKo02}.

In this Letter we show that yet there exists another microscopic mechanism
creating non-monotonicities in the frequency-dependent diffusion
coefficient. Our starting point is to study diffusion in the BBB at smaller
friction and smaller scatterer radius than in Ref.\ \cite{MaKl03}. In this
situation we find a {\em spiraling bouncing ball mode} that changes under
parameter variation. We show that this mode is again responsible for the
emergence of local maxima and minima in the diffusion coefficient as a
function of the vibration frequency. We conclude with a brief outlook towards
consequences of our work for the standard bouncing ball as well as for
transport on vibro-transporters.

\section{The bouncing ball billiard}

\begin{figure}
\includegraphics[height=14cm,angle=-90]{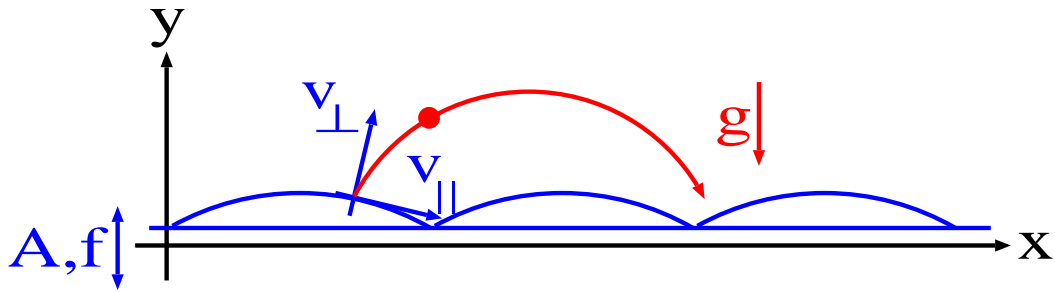}
\caption{Sketch of the bouncing ball billiard: A point particle
collides inelastically with circular scatterers forming a periodic lattice on
a line. Parallel to $y$ there is an external field with constant acceleration
$g$. The corrugated floor oscillates with amplitude $A$ and frequency $f$.}
\label{fig:bbb}
\end{figure}

The model is depicted in Fig.\ \ref{fig:bbb}: It consists of a floor
oscillating with $y_f=-A\,\sin(2\pi f\, t)$, where $A=0.01$ and $f$ are the
amplitude respectively the frequency of the vibration.\footnote{We work with
reduced units, where all quantities are dimensionless in terms of the unit
amplitude $A_0=1mm$, respectively the unit frequency $f_0=1$Hz.}  This floor
is equipped with a periodic grid of circular scatterers of radius $R$ whose
centers are a distance of $S=2$ apart from each other. We now consider a point
particle\footnote{Note that the radii of the moving particle and of a
scatterer are additive, hence we only neglect any rotational energy.}
performing a free flight between two collisions in a gravitational field with
acceleration $g=9800\parallel y$. The particle's coordinates
$(x^-_{n+1},y^-_{n+1})$ and velocities $(v^-_{x,n+1},v^-_{y,n+1})$ at time
$t_{n+1}$ immediately before the $(n+1)$th collision and its coordinates
$(x^+_n,y^+_n)$ and velocities $(v^+_{x\,n},v^+_{y\,n})$ at time $t_n$
immediately after the $n$th collision are related by the equations
\cite{MaKl03} 
\begin{eqnarray} 
x^-_{n+1} &=& x^+_{n} + v^+_{x,n}(t_{n+1}-t_{n})  \label{eq:eomx}\\ 
y^-_{n+1} &=& y^+_{n} + v^+_{y,n} (t_{n+1}-t_{n}) - g(t_{n+1}-t_{n})^2/2 \label{eq:eomy}\\
v^-_{x,n+1}&=& v^+_{x,n}  \\ 
v^-_{y,n+1} &=& v^+_{y,n} - g (t_{n+1}-t_{n})   \label{eq:eomvy}\:.
\end{eqnarray}
At a collision the velocities change according to
\begin{eqnarray}
v^+_{\perp,n}-v_{f\perp,n} &=& \alpha \,(v_{f\perp, n}-v^-_{\perp, n}) \label{eq:collperp}\\
v^+_{\parallel , n}-v_{f\parallel , n} &=& \beta \,(v^-_{\parallel
,n}-v_{f\parallel , n}) \:,  \label{eq:collpar}
\end{eqnarray} 
where $v_f$ is the velocity of the corrugated floor. The local coordinate
system for the velocities $v_{\perp}$ and $v_{\parallel}$ is given in Fig.\
\ref{fig:bbb}. We assume that the scattering process is inelastic by
introducing the two restitution coefficients $\alpha$ and $\beta$
perpendicular, respectively tangential to the surface at the scattering point.

In Ref.\ \cite{MaKl03} we studied the BBB for $R=25$, $\alpha=0.5$ and
$\beta=0.99$. Here we consider the case of $R=15$ and $\alpha=0.7$, which is
closer to the experiments of Urbach et al.\ \cite{PrEgUr02}. In contrast to
naive expectations, this simple variation of parameters profoundly changes the
diffusive dynamics of the BBB, as we will show in the following.

\section{Deterministic diffusion, frequency locking, and spiral modes}

The diffusion coefficient $D$ was computed from simulations according to the
Einstein formula
\begin{figure}
\includegraphics[width=14cm]{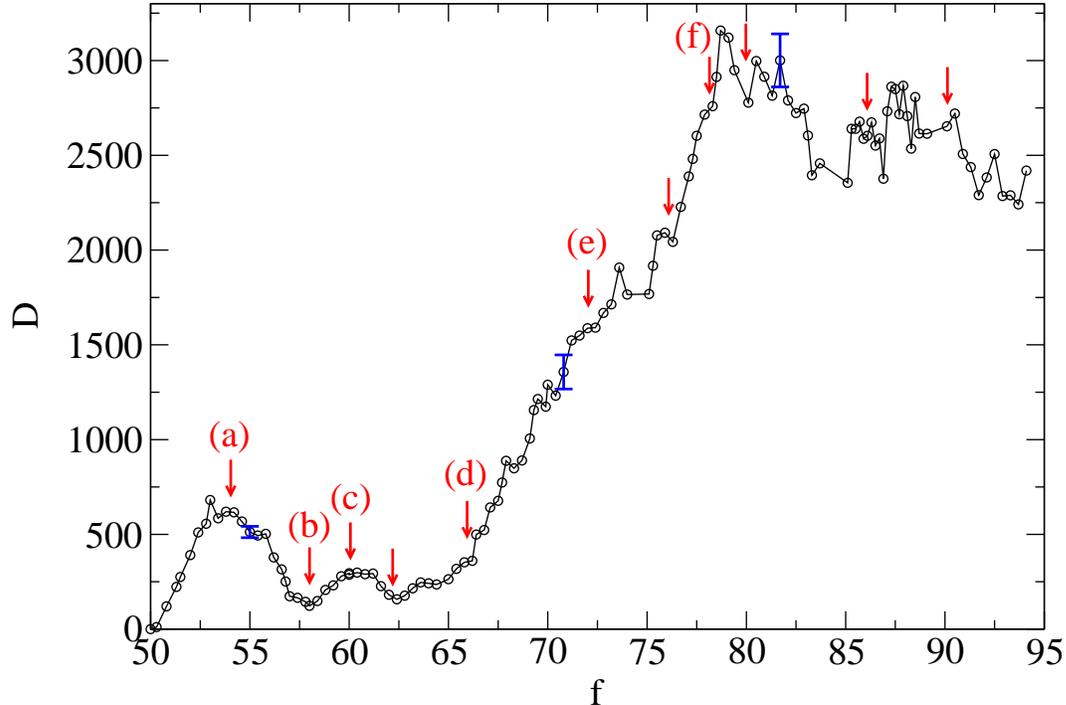}
\caption{The diffusion coefficient $D(f)$ of the bouncing
ball billiard Fig.\ \ref{fig:bbb} at friction $\alpha=0.7$ and scatterer
radius $R=15$ as a function of the vibration frequency $f$ of the corrugated
floor. The graph consists of $124$ data points with error bars at $f=55, 70.8$
and $82.1$. The labels (a) to (e) refer to the respective phase space plots of
Fig.\ \ref{fig:proj}.}
\label{fig:df}
\end{figure}

\begin{figure}
\includegraphics[height=20cm]{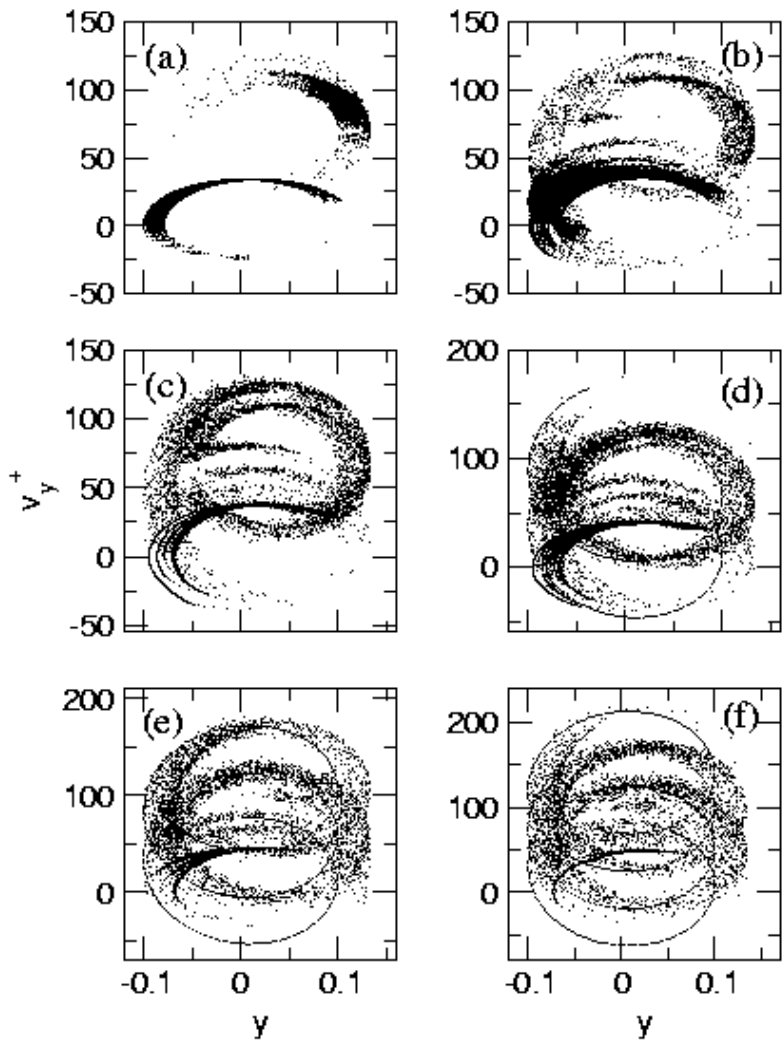}
\caption{Projections of the phase space of the bouncing ball billiard Eqs.\
(\ref{eq:eomx}) to (\ref{eq:collpar}) onto positions and velocities along the
$y$-axis at the collisions. The frequencies of the vibrating plate are $f=$
(a) $54$, (b) $58$, (c) $60$, (d) $66$, (e) $72$, (f) $78$. The spiraling
lines represent the analytical approximation Eqs.\
(\ref{eq:appy}),(\ref{eq:appvy}).}
\label{fig:proj}
\end{figure}
\begin{equation} 
D = \lim_{t\rightarrow \infty} \frac{<(x(t)-x(0))^2>}{2t}\:,
\label{eq:einstein} 
\end{equation} 
where the brackets denote an ensemble average over moving particles; for
further numerical details cf.\ Ref.\ \cite{MaKl03}. Fig.\ \ref{fig:df} shows
that $D$ exhibits local maxima and minima as a function of the driving
frequency $f$. However, already at a first view this curve looks very
different from the corresponding one in Ref.\ \cite{MaKl03}, where the
diffusion coefficient was computed at different parameter values.

The major irregularities of Fig.\ \ref{fig:df} can be understood by means of
the phase space projections displayed in Fig.\ \ref{fig:proj}. Here we have
plotted the positions $y$ and the velocities $v_y^+$ of a moving particle
immediately after its collisions with the plate. The onset of diffusion around
$f=50$ is characterized by the existence of many {\em creeping orbits}, where
the particle performs long sequences of tiny little jumps along the surface
hence moving parallel to $y$ like the harmonically oscillating plate. This
dynamics is reminiscent in Fig.\ \ref{fig:proj} in form of incomplete ellipses
around $v_y^+=0$ thus representing a simple harmonic oscillator mode.

The first maximum of $D(f)$ around $f=54$ reflects a $1/1$ frequency locking
between particle and plate, as is shown in Fig.\ \ref{fig:proj} (a) as the
spot around $v_y^+\simeq100$. In other words, the time of flight $T_p$ of the
particle is identical with the vibration period $T_f$ of the floor.
Additionally, the system may exhibit a dynamics where this behavior is
interrupted by periods of creepy motion. Whether this happens depends on the
initial conditions, hence the dynamics is non-ergodic. However, only for the
peak around $f=55$ we could detect such difficulties, whereas for $f\ge58$ our
numerics indicated ergodic motion. Frequency locking has been
widely discussed for the simple bouncing ball
\cite{Pier85,KFP88,TAR92,LuMe93,WCBH96,HCF89,HoFi89} and has already been
found to enhance diffusion in the BBB \cite{MaKl03}, where all the major peaks
of the frequency-dependent diffusion coefficient could be identified in terms
of integer frequency locking.

Around $f=58$ this resonance is completely destroyed resulting in a local
minimum of $D(f)$. In Fig.\ \ref{fig:proj} (b) this is represented by a
``smearing out'' of the $1/1$ frequency locking spot together with a dominance
of creeps. Surprisingly, the next local maximum at $f=60$ corresponds to a new
structure in phase space: The granular particle locks into a {\em virtual harmonic
oscillator mode} (VHO) situated around $v_y^+\simeq 70$, cf.\ Fig.\
\ref{fig:proj} (c), which reflects the harmonic driving and again enhances
diffusion. Around $f=62$ this mode again becomes unstable leading to a local
minimum. At $f=66$ the VHO starts to ``spiral out'', see Fig.\
\ref{fig:proj} (d), eventually resulting in a two-loop spiral, Fig.\
\ref{fig:proj} (e). This causes a drastic increase of $D(f)$ on a global
scale. Note that we could not detect any simple periodic motion on this
spiral, hence this scenario may not be understood as a simple period-doubling
bifurcation. We furthermore emphasize that parallel to $x$ the motion is
always highly irregular, as is reminiscent in the existence of a diffusion
coefficient. At $f=76$ another loop develops, see Fig.\
\ref{fig:proj} (f) with $f=78$ for a slightly advanced stage, before this
structure settles into a three-loop spiral that explains the washed-out local
maximum in $D(f)$ around $f=80$. A fourth loop emerges at $f=86$ being
completed around $f=90$, followed by the onset of a fifth one at $f=96$.

In order to understand the frequency dependence of this spiral more
quantitatively we analyze the data presented in Fig.\ \ref{fig:proj} using a
simple argument of Warr et al.\ \cite{WCBH96}: The time of flight $T_p$
between two collisions of the particle with the plate is determined by
$T_p=2v_y^+/g$, hence one may calculate
\begin{equation}
k:=\frac{T_p}{T_f}=\frac{2v_y^+f}{g}
\end{equation}
and check whether this fraction yields some rational number indicating some
frequency locking. Extracting $v_y^+$ from Fig.\ \ref{fig:proj} one can easily
verify that the black spot in (a) indeed corresponds to $k\simeq1$. Complete
spiral loops at higher frequencies appear to be situated around $f_c+c\;\Delta
f$, $c\in\mathbb{N}_0$, with $f_c\simeq 60$ and $\Delta f\simeq 10$. Let
$v_y^+$ be the maximal values of the spirals in Fig.\ \ref{fig:proj}. We then
find that $k\simeq1.47$ around $f=60$, $k\simeq2.5$ around $f=72$,
$k\simeq3.43$ around $f=80$, and $k\simeq4.5$ around $f=90$. The spiral mode
thus locks into ``smeared-out'' half-integer resonances starting around
$f_i+i\;\Delta f$, $i\in\mathbb{N}_0$, $f_i\simeq 66$. 

We remark that, for a ball bouncing on a flat plate, signs of such spiral
modes were already observed experimentally \cite{WCBH96} and described
theoretically \cite{KFP88,TAR92,LuMe93,GiMa03}. Particularly, Luck and Mehta
\cite{LuMe93} provided a simple analytical approximation for the coarse
functional form of this mode. Their derivation can straightforwardly be
repeated for Eqs.\ (\ref{eq:eomx}) to (\ref{eq:collpar}): Let us approximate
the surface to be flat, and let us assume that there are no correlations
between the collisions. By using Eqs.\ (\ref{eq:eomy}) and (\ref{eq:collperp})
we then arrive at
\begin{eqnarray} 
y_1^+&=&-A\sin(2\pi ft_1) \label{eq:appy}\\ 
v_{y,1}^+&=&\alpha g/2(t_1-t_0)-A2\pi f(1+\alpha)\cos(2\pi f t_1)
\label{eq:appvy}\;,
\end{eqnarray}
cp.\ Eq.\ (\ref{eq:appvy}) to Eq.\ (3.20) of Ref.\ \cite{LuMe93}, where $t_0$
is the initial time at which the particle was launched from the plate and
$t_1$ is the time of the first collision. For $t_0$ we calculated the time at
which the plate moves with maximum positive velocity, since here, for high
enough frequency, a particle will be launched which previously is at rest. For
$t_1$ we obtained the distributions of collision times $t_c$ from simulations
yielding a range of $0\le t_c\le t_{max}(f)$. In order to ensure that
$t_1-t_0\ge 0$ we then defined $t_1:=t_c+t_0$. Consequently, Eqs.\
(\ref{eq:appy}), (\ref{eq:appvy}) contain no fit parameter. Results
from Eqs.\ (\ref{eq:appy}), (\ref{eq:appvy}) are displayed in Fig.\
\ref{fig:proj} (d) to (f). Naturally, this approximation does not work well for
small collision times, respectively in the regime of creeping orbits. However,
starting from the VHO it reproduces the whole structure of the spiral mode
very well.

\section{Outlook and conclusions}

In this Letter we have studied the BBB for smaller friction and smaller
scatterer radius than in our previous work Ref.\ \cite{MaKl03}. Computer
simulations demonstrated that, again, the diffusion coefficient is a highly
irregular function of the vibration frequency. However, for the present
parameters only the first local maximum could be identified in terms of a
simple frequency locking. We showed that there exist local extrema due to what
we called a spiral mode in the BBB. Whenever the vertical velocity of the
particle, respectively the vibration frequency of the plate, is large enough
such that the particle can lock into a multiple half-integer frequency of the
plate, the spiral mode develops another loop. The coarse functional form of
this spiral was extracted from the equations of motion by neglecting the
radius of the scatterers and by assuming memory loss at the collisions.

In Ref.\ \cite{GiMa03} these spiral structures have been denoted as
``chattering bands'' of the simple bouncing ball, and it has been argued that
they are rather stable against random perturbations. In the BBB the scatterers
are defocusing inducing an additional deterministic chaotic component into the
dynamics \cite{KlKo02}. Interpreting the geometry of the BBB as a
``perturbation'' thus leads to the conclusion that spiral modes form states of
the bouncing ball dynamics which are much more stable than simple frequency
locking resonances.

Interestingly, bouncing ball-type models have already been used since quite
some time in order to understand transport on vibro-transporters, which are
common carriers of agricultural material and related matter. For a corrugated
version that very much reminds of the BBB and is widely used in industrial
applications see, e.g., Ref.\ \cite{PeMe92}. In order to transport particles
on such devices one generates an average drift velocity by means of some
symmetry breaking. For tilted systems with a flat surface Hongler et al.\
\cite{HCF89,HoFi89} predicted an integer frequency locking leading to local
extrema in the transport rate as a function of a control parameter, as was
recently verified in experiments \cite{HaLe02}.  Grochowski et al.\
constructed a circular vibratory conveyor, where the trough was driven by
asymmetric oscillations \cite{GSWK03,RKRGW03,GWRKR03}.  Both for a tracer
particle in a layer of granular matter and for a single highly inelastic test
particle they observed a non-monotonic current reversal of the transport
velocity under variation of control parameters. This ratchet-like effect was
recently reproduced in simulations \cite{ElLi03} again indicating some
underlying frequency locking.

Our work leads to the conclusion that in vibro-transporters not only the
current, but also other transport property quantifiers such as the diffusion
coefficient may be irregular functions of control parameters. So far we have
identified three basic mechanisms generating irregular parameter dependencies,
which are (i) frequency locking \cite{MaKl03}, (ii) the spiral modes discussed
here, and (iii) pruning effects \cite{KlKo02} leading to fractal-like
irregularities on finer and finer scales. Looking for irregular transport
coefficients in experiments may thus pave the way for a very fine-tuned
control of the dynamics of vibro-transporters. This may eventually provide a
basis for practical applications such as, for example, sophisticated methods
of particle separation \cite{FSWV02}.

\begin{ack}
R.K.\ thanks Chr.\ Kr\"ulle, J.\ Urbach, H.\ Elhor and A.\ Riegert for helpful
correspondence and interesting discussions. L.M.\ was supported by the
European Community under the contract No. HPMF-CT-2002-01511.
\end{ack}


\begin{thebibliography}{10}
\expandafter\ifx\csname url\endcsname\relax
  \def\url#1{\texttt{#1}}\fi
\expandafter\ifx\csname urlprefix\endcsname\relax\def\urlprefix{URL }\fi

\bibitem{Pier83}
P.~Pieranski, Jumping particle model. {P}eriod doubling cascade in an
  experimental system, J. Physique 44 (1983) 573--578.

\bibitem{Pier85}
P.~Pieranski, Z.~Kowalik, M.Franaszek, Jumping particle model. {A} study of the
  phase space of a non-linear dynamical system below its transition to chaos,
  J. Physique 46 (1985) 681--686.

\bibitem{KFP88}
Z.~Kovalik, M.~Franaszek, P.~Pieranski, Self-reanimating chaos in the
  bouncing-ball system, Phys. Rev. A 37 (1988) 4016--4022.

\bibitem{TAR92}
N.~Tufillaro, T.~Abbott, J.~Reilly, An experimental approach to nonlinear
  dynamics and chaos, Add. Wesley Publ., New York, 1992.

\bibitem{LuMe93}
J.~Luck, A.~Mehta, Bouncing ball with a finite restitution: Chattering,
  locking, and chaos, Phys. Rev. E 48 (1993) 3988--3997.

\bibitem{WCBH96}
S.~Warr, W.~Cooke, R.~Ball, J.~Huntley, Probability distribution functions for
  a single particle vibrating in one dimension: Experimental study and
  theoretical analysis, Physica A 231 (1996) 551--574.

\bibitem{GiMa03}
S.~Giusepponi, F.~Marchesoni, The chattering dynamics of an ideal bouncing
  ball, Europhys. Lett. 64 (2003) 36--42.

\bibitem{LCG99}
W.~Losert, D.~Cooper, J.~Gollub, Propagating front in an excited granular
  layer, Phys. Rev. E 59 (1999) 5855--5861.

\bibitem{FSWV02}
Z.~Farkas, F.~Szalai, D.~Wolf, T.~Vicsek, Segregation of granular binary
  mixtures by a ratchet mechanism, Phys. Rev. E 65 (2002) 022301/1--4.

\bibitem{PrEgUr02}
A.~Prevost, D.~Egolf, J.~Urbach, Forcing and velocity correlations in a
  vibrated granular monolayer, Phys. Rev. Lett. 89 (2002) 084301/1--4.

\bibitem{MaKl03}
L.~M\'aty\'as, R.~Klages, Irregular diffusion in the bouncing ball billiard,
  Physica D 187 (2004) 165--183.

\bibitem{HCF89}
M.-O. Hongler, P.~Carter, P.~Flury, Numerical study of a model of
  vibro-transporter, Phys. Lett. A 135 (1989) 106--112.

\bibitem{HoFi89}
M.-O. Hongler, J.~Figour, Periodic versus chaotic dynamics in vibratory
  feeders, Helv. Phys. Acta 62 (1989) 68--81.

\bibitem{KlKo02}
R.~Klages, N.~Korabel, Understanding deterministic diffusion by correlated
  random walks, J. Phys. A: Math. Gen. 35 (2002) 4823--4836.

\bibitem{PeMe92}
S.~Persson, M.~Megnin, Factors affecting material movement on a corrugated
  vibrating conveyor (grainpan), Trans. ASAE 35 (1992) 395--400.

\bibitem{HaLe02}
I.~Han, Y.~Lee, Chaotic dynamics of repeated impacts in vibratory bowl feeders,
  J. Sound. Vibr. 249 (2001) 529--541.

\bibitem{GSWK03}
R.~Grochowski, S.~Strugholtz, P.~Walzel, C.~Kr\"ulle, Granular transport on
  vibratory conveyor, Chemie Ingenieur Technik 75 (2003) 1103.

\bibitem{RKRGW03}
M.~Rouijaa, C.~Kr\"ulle, I.~Rehberg, R.~Grochowski, P.~Walzel,
  Trans\-port\-verhalten und {S}trukturbildung granularer {M}aterie auf
  {S}chwingf\"orderern, Chemie Ingenieur Technik 76 (2004) 62--65.

\bibitem{GWRKR03}
R.~Grochowski, P.~Walzel, M.~Rouijaa, C.~Kr\"ulle, I.~Rehberg, Reversing a
  granular flow on a vibratory conveyor, Appl. Phys. Lett. 84 (2004)
  1019--1022.

\bibitem{ElLi03}
H.~Elhor, S.~Linz, Calculation of transport velocity of granular material on a
  vibratory conveyer, poster contribution to the conference {\em Dynamik auf
  Netzwerken}, MPIPKS Dresden (Nov.\ 2003).

\end{thebibliography}

\end{document}